# High-Speed Optical Microscopy for Neural Voltage Imaging: Methods, Trade-offs, and Opportunities

Zhaoqiang Wang[1], Ruth R. Sims[2], Sheng Xiao[3], Ruixuan Zhao[1], Ohr Benshlomo[1], Zihan Zang[1], Jiamin Wu[4], Valentina Emiliani[2], Liang Gao[1, 5*]

[1]Department of Bioengineering, University of California, Los Angeles, CA, USA 90095

[2] Institut de la Vision, Sorbonne Université, INSERM, CNRS, Paris, France 75012

[3]Chan Zuckerberg Biohub, San Francisco, CA, USA 94158

[4]Department of Automation. Tsinghua University, China 100084

[5]California NanoSystems Institute, University of California, Los Angeles, CA, USA 90095

[*]Correspondence: gaol@ucla.edu

## ABSTRACT

High-speed optical imaging of dynamic neuronal activity is essential yet challenging in neuroscience. While calcium imaging has been firmly established as a workhorse technique for monitoring neuronal activity, its limited temporal resolution and indirect measurement restrict its ability to capture rapid inhibitory and excitatory events and subthreshold voltage oscillations. In contrast, voltage imaging directly measures membrane potential fluctuations, providing a comprehensive and precise representation of neuronal circuit dynamics. Recent advancements in voltage-sensitive dyes and genetically encoded voltage indicators have significantly enhanced the feasibility of voltage imaging, prompting the development of advanced fluorescence microscopy methods optimized for high-speed acquisition. However, achieving millisecond-scale temporal resolution remains challenging due to inherent trade-offs among imaging speed, spatial resolution, and signal-to-noise ratio. Conventional raster-scanning approaches, including confocal microscopy, are fundamentally limited by their slow frame rates, precluding the capture of rapid neuronal events from multiple neurons simultaneously. Alternative techniques such as random-access scanning, spatiotemporal multiplexing, and computational optical imaging have successfully addressed these constraints, enabling kilohertz-level imaging of neuronal activity in both two-dimensional and three-dimensional contexts. This review summarizes recent progress in high-speed optical microscopy for voltage imaging and discusses its transformative potential for neuroscience research.

## INTRODUCTION

Communication between neurons is mediated by rapid fluctuations in membrane potential, and understanding neural circuits therefore requires tools that can record activity across large neuronal populations *in vivo*. Optical imaging offers key advantages over electrode-based approaches by enabling minimally invasive, simultaneous recordings from thousands of neurons, high-resolution spatial mapping of activity, and access to genetically defined cell classes. Optical measurements can also capture subthreshold membrane potential dynamics that are difficult to access with population electrophysiology, reduce bias toward highly active neurons, and facilitate monitoring of neuronal subcompartments that remain challenging for patch-clamp recordings. Together, these capabilities reveal rich spatiotemporal activity patterns—such as propagating signals across circuits[1–6]—that are often difficult to reconstruct with conventional electrophysiology.

Among optical approaches, calcium imaging has become a widely used workhorse for monitoring population neural activity and has enabled numerous foundational discoveries in neuroscience.[7–9] However, calcium signals are only an indirect proxy for electrical activity.[10,11] In most calcium-imaging experiments, the recorded signal reflects the response of a calcium indicator to underlying changes in intracellular calcium concentration, which may arise from calcium influx through voltage-gated calcium channels, ionotropic glutamate receptors, and related pathways, as well as from release from intracellular stores. Although calcium entry itself can be rapid in small compartments, the indicator-reported calcium signal typically evolves on timescales slower than the underlying membrane potential dynamics. As a result, calcium imaging is poorly suited for resolving precise spike timing. Moreover, it is largely insensitive to subthreshold voltage events—including excitatory and inhibitory postsynaptic potentials—and cannot report hyperpolarizing signals, rendering key aspects of synaptic integration, inhibition, and plasticity inaccessible.[12–14] In contrast, voltage imaging directly reports membrane potential with millisecond temporal precision, providing a more faithful and comprehensive depiction of neuronal activity.[1,3,15–17]

Despite this promise, voltage imaging imposes stringent technical demands. During an action potential, the electric field reorients within one to two milliseconds, with the fastest component—the rising phase—occurring in approximately 250 µs (Box 1). This makes action potentials 100–1000-fold faster than calcium transients, necessitating high imaging speeds to accurately capture their dynamics.[18] In practice, voltage imaging is typically performed at frame rates over 300 Hz[14] and preferably in the kilohertz range[18,19] to avoid waveform distortion or missed events.

Such acquisition rates severely limit photon collection time, making voltage imaging intrinsically noisier than calcium imaging. Increasing excitation intensity can partially compensate for reduced exposure time, but at the cost of faster photobleaching, shorter recording durations, and increased risk of photodamage due to heating or nonlinear effects. In addition, voltage signals originate from the plasma membrane, where the number of contributing fluorophores is inherently limited, further constraining signal strength. Finally, imperfect membrane targeting of voltage indicators can increase background fluorescence from non-responsive molecules, elevating shot noise and degrading both the signal-to-background ratio (SBR) and signal-to-noise ratio (SNR).

On the detection side, maintaining a large—or three-dimensional (3D)—field of view (FOV) while resolving fine neuronal features remains challenging because kilohertz imaging demands immense pixel throughput. Sustaining these data rates over extended recordings further stresses data transmission bandwidth, real-time throughput, and storage capacity. In this review, we highlight recent advances in optical microscopy that help overcome these constraints and enable high-speed one-photon (1P) and two-photon (2P) voltage imaging. Collectively, these developments expand our ability to capture rapid neuronal dynamics with high spatiotemporal fidelity, opening new opportunities to probe cooperative signaling and large-scale organization in neural networks.

## 2. STATE-OF-THE-ART ONE-PHOTON VOLTAGE IMAGING

Currently, 1P excitation microscopy remains the dominant platform for voltage imaging because of its high excitation efficiency and compatibility with high-throughput widefield detection. In this section, we review state-of-the-art 1P voltage imaging methods and organize them into three broad categories: (*i*) planar approaches that maximize temporal throughput (*e.g.*, widefield and targeted illumination), (*ii*) multi-depth strategies that provide optical sectioning and volumetric access (*e.g.*, multi-Z confocal and light-sheet microscopy), and (*iii*) snapshot light-field approaches that reconstruct 3D activity at camera-limited volumetric rates. Representative performance metrics are summarized in Table 1.

### 2.1. PLANAR VOLTAGE IMAGING

Conventional 1P voltage imaging has often relied on widefield microscopy, which captures a two-dimensional (2D) FOV in a single snapshot. This intrinsic detection parallelism provides a high-throughput, low-cost, and experimentally straightforward path to the high frame rates required for resolving fast voltage dynamics. A key limitation, however, is the lack of optical sectioning: out-of-focus fluorescence contributes substantially to the detected signal, elevating background and reducing the SBR (Box 2), particularly *in vivo*. To address this challenge, recent work has combined widefield imaging with molecular strategies—including soma targeting[1,20–22] and sparse labeling[1,23]—to reduce diffuse background and crosstalk. Complementary optical approaches further suppress background by restricting excitation to relevant structures, using structured[24,25], holographic[26,27] or targeted illumination[1,3,16,28,29].

Despite these advances, widefield microscopy has so far been used primarily at shallow depths and in sparsely labeled preparations. Imaging densely labeled populations and deeper brain regions instead favors optical-sectioning modalities that physically suppress out-of-focus fluorescence. However, conventional optical-sectioning approaches such as confocal microscopy typically rely on point scanning, which intrinsically limits frame rate—particularly over large FOVs—making it challenging to capture fast voltage dynamics across large neuronal populations.

To increase throughput, line-scan confocal microscopy has been adopted for voltage imaging, reducing the scanning requirement to one dimension by illuminating the sample with a focused line and detecting the resulting fluorescence with a linear detector array.[30] Although line scanning sacrifices some background rejection relative to point-scanned confocal microscopy, it enables kilohertz imaging over millimeter-scale FOVs. In addition, the reduced demands on scan speed permit the use of large-aperture scanners, facilitating efficient fluorescence collection at high numerical aperture (NA). To offset the diminished background suppression, Xiao et al. combined line-scan confocal microscopy with targeted illumination, selectively exciting only neurons of interest (Fig. 1a).[30] This strategy achieves over a 50-fold enhancement in SBR, extending the imaging depth to 300 μm, sufficient to image neurons in cortical layer 3 *in vivo*.

Wide-field fluorescence lifetime imaging also represents a promising new direction for planar voltage imaging. Using electro-optic fluorescence lifetime imaging microscopy (EO-FLIM), Bowman et al. achieved kilohertz-rate wide-field lifetime imaging of neuronal voltage signals *in vivo*, resolving both action potential propagation and subthreshold activity in adult *Drosophila*.[31] Because lifetime readout is less sensitive to photobleaching, this work highlights fluorescence lifetime as a potentially valuable complement to conventional intensity-based approaches.

## 2.2. MULTI-DEPTH VOLTAGE IMAGING

To extend voltage imaging to 3D, Multi-Z confocal microscopy[32] employs axially extended illumination (via a lower NA) and a cascade of reflective pinholes in the detection path to acquire signals from multiple depths simultaneously (Fig. 1b). Fluorescence rejected by one pinhole is relayed to subsequent pinholes, minimizing signal loss and maintaining high overall detection efficiency. A recent variant, multi-Z imaging with confocal detection (MuZIC)[33], incorporates an ultrafast polygon scanner to achieve a 117 kHz line rate—approximately an order of magnitude faster than resonant galvanometric scanners. Combined with a second (linear) galvo and axially distributed pinholes, MuZIC simultaneously scans four imaging planes (128 × 127 pixels per plane) at 916 Hz. This performance enables recording of subthreshold and spiking dynamics in mouse motor cortex expressing Voltron2-ST[28] through a cranial window.

Further acceleration of Multi-Z confocal imaging can be achieved by adopting line scanning rather than point scanning.[34] Tsang et al. replaced pinholes with reflective slits and demonstrated simultaneous imaging of three planes, enabling voltage recordings of FluoVolt-labeled cardiomyocytes across a 575 × 645 × 64 $\mu m^3$ volume with clearly resolved action potential waveforms. Notably, multiplane imaging also offers robustness to axial brain motion, as signals in adjacent planes help distinguish true membrane voltage fluctuations from motion-induced artifacts—an important advantage over conventional 2D imaging, particularly *in vivo*.

Additional gains in imaging speed can be achieved by further reducing scan dimensionality. Instead of scanning a point or a line, light-sheet fluorescence microscopy (LSFM) acquires images plane-by-plane: a thin sheet of light selectively illuminates a single optical section, and a 2D camera records that plane in a single exposure (Fig. 2a). This selective planar excitation provides optical sectioning while minimizing phototoxicity, and the use of widefield detection offers advantages over confocal microscopy in terms of simpler implementation, higher light efficiency, and faster acquisition.[35,36,37]

High-resolution voltage imaging with LSFM typically requires a high-NA detection objective, which comes with a shallow depth of field. Consequently, during 3D imaging the detection focal plane must be translated in synchrony with the axially scanned light sheet to maintain focus on the illuminated section. A common implementation physically moves the imaging objective using a piezoelectric actuator; however, the inertia of the objective limits the speed and efficiency of multi-plane acquisition. Remote focusing offers an attractive alternative for fast volumetric imaging: rather than translating the primary objective, an additional refocusing element is inserted into the detection path to rapidly shift the focal plane. This can be achieved by introducing controlled defocus using a tunable lens in the Fourier plane or a piezo-driven mirror in a conjugate image plane.[38–40] The mirror-based approach is often preferred, as it can provide aberration-free refocusing over a relatively large axial range. This strategy was recently demonstrated for whole-brain voltage imaging in larval zebrafish.[41]

Mirror-based remote-focusing LSFM requires the incident rays to be nominally normal to the refocusing mirror in order to introduce controlled defocus. This is commonly implemented using a beamsplitter to co-align the incident and reflected paths, but the beamsplitter incurs substantial light loss—an acute limitation for voltage imaging, where exposure times are short and SNR is paramount. To overcome this inefficiency, Böhm et al. introduced flipped image remote focusing (FLIPR), which replaces the beamsplitter with a microscopic retroreflector (Fig. 2b).[42] The retroreflector flips and folds the image back into the remote-focusing arm, approximately doubling light efficiency relative to conventional beamsplitter-based designs that use a polarization beamsplitter together with a quarter-wave plate.

The volumetric frame rate of remote-focusing LSFM is ultimately limited by the focal switching speed of the refocusing elements. To circumvent this constraint, Sacconi et al. introduced a parallelized detection strategy that uses multiple cameras to acquire seven planes simultaneously.[43] In their implementation, several light sheets are projected into the sample in parallel, and fluorescence from each illuminated plane is directed to a dedicated camera. This architecture enabled scan-free 3D imaging at kilohertz rates and was demonstrated by recording the propagation of electrical impulses in voltage-dye–labeled zebrafish hearts. However, scaling camera-array approaches remains challenging: repeated beam splitting reduces the photon budget per plane (and thus SNR), and exciting multiple planes simultaneously increases out-of-focus background, further degrading contrast.

## 2.3. SNAPSHOT VOLUMETRIC VOLTAGE IMAGING

Light-field microscopy (LFM) is a leading snapshot 3D imaging modality and has been successfully applied to voltage imaging. Unlike conventional widefield microscopy, LFM captures both spatial and angular information using an angular-demultiplexing element such as a microlens array. The resulting 4D dataset ($x, y, \theta, \phi$)—where ($x, y$) denote spatial coordinates and ($\theta, \phi$) represent emission angles—can be computationally refocused to reconstruct volumetric images by leveraging disparities among angular views.[44–47] Importantly, because the full light field is recorded, optical aberrations can be estimated and corrected in postprocessing by digitally manipulating the captured rays (digital adaptive optics), eliminating the need for deformable mirrors or spatial light modulators.[48,49] This capability enables multi-site aberration correction without sacrificing acquisition speed, which is particularly valuable for high-fidelity intravital imaging in heterogeneous environments and across mesoscale FOVs.[50] Since LFM is scanless, its volumetric imaging rate is primarily determined by camera readout speed. For example, using 1P LFM, Quicke et al. imaged action potential–evoked fluorescence transients in mouse brain slices sparsely expressing the GEVI VSFP-Butterfly 1.2.[51] At 100 volumes per second, they demonstrated simultaneous capture of axially separated dendrites and single-shot localization of GEVI transients within a 3D dendritic arbor.

Although LFM removes the constraints of mechanical scanning, volumetric rate is often limited by camera electronic bandwidth. Recording a full 4D light-field dataset typically requires large-format sensors, which frequently operate at reduced frame rates, making kilohertz volumetric imaging challenging. To address this "big data" bottleneck, Guo et al. developed a neuromorphic LFM framework that replaces the conventional 2D sensor with an event camera.[52] Rather than recording full frames, event cameras capture only pixel-wise intensity changes, substantially reducing data volume and better utilizing bandwidth to enable kilohertz rates. However, current event cameras remain limited for voltage imaging by several practical factors, including lower pixel counts than modern scientific CMOS cameras, finite contrast sensitivity thresholds, and reduced performance for small-amplitude analog signals near the detection threshold. As a result, although they are well suited for capturing sparse, fast transients, they are currently less effective for faithfully recording weak continuous voltage fluctuations such as subthreshold oscillations. These limitations are not entirely fundamental: ongoing improvements in pixel density, back-side illumination, fill factor, contrast sensitivity, and readout architecture should further expand their utility for voltage imaging.

A complementary strategy to reduce data burden is compressive light-field acquisition. Wang et al. introduced squeezed light-field microscopy (SLIM),[53] motivated by the observation that perspective images in conventional LFM are highly redundant, differing primarily by depth-dependent disparities across subaperture views.[54] SLIM exploits this redundancy by compressing perspective measurements along selected orientations and redistributing them across views, greatly reducing dataset size while enabling high-resolution 3D imaging on a low-pixel-count, high-speed camera. Using this approach, Wang et al. achieved kilohertz 3D imaging across multiple biological systems, including *in vivo* recordings of action potentials and subthreshold oscillations in mice (Fig. 3a).

Beyond data throughput, LFM faces fundamental challenges in spatial resolution. Because LFM partitions the objective pupil across angular views, it inherently trades diffraction-limited lateral resolution for depth information. Several approaches have been proposed to recover resolution. Scanning light-field microscopy (sLFM) reintroduces scanning by translating the intermediate image across the microlens array, leveraging the microlens-induced diffraction structure to recover high-frequency content and approach diffraction-limited resolution using a temporal sliding window.[48] While effective, the required scanning can introduce motion artifacts in highly dynamic samples. To avoid physical scanning, Lu et al. developed Virtual-Scanning Light Field Microscopy (VsLFM), which replaces mechanical scanning with a physics-guided deep-learning reconstruction.[55] By exploiting phase correlations across angular views, their virtual-scanning network (Vs-Net) achieved a near diffraction-limited resolution, while retaining scanless acquisition at the camera's native frame rate (Fig. 3b).

The speed advantage of LFM comes with important trade-offs. LFM is inherently tomographic, acquiring information through nonuniform sampling in spatial-frequency space that typically oversamples low frequencies while undersampling high frequencies. This makes it intrinsically less efficient than direct spatial-domain scanning for uniform, high-fidelity recovery of fine details. Moreover, because tomographic measurements are multiplexed, signals from different locations are mixed on the detector, reducing the effective dynamic range. If bright and dim objects are present in the same FOV, shot noise from the bright object can propagate through the reconstruction and contaminate the dim object, and this effect cannot be removed numerically after detection. More broadly, like other point-spread-function-engineering approaches, LFM encodes 3D position into a 2D camera pattern by spreading photons over many pixels. While reconstruction can computationally reassign these signals to their object locations, the accumulated detector noise is also propagated, ultimately reducing SNR.

Finally, unlike confocal microscopy and LSFM, LFM lacks intrinsic optical sectioning. Although computational refocusing can reconstruct depth planes, out-of-focus fluorescence cannot be physically rejected, which reduces contrast—especially in deeper tissue or when background consumes a large fraction of the camera dynamic range. To address this, Bai et al. adopted confocal light-field microscopy (confocal LFM), which integrates confocal line detection with LFM to suppress background fluorescence.[56,57] In confocal LFM, a laser line scans across the FOV, and the resulting fluorescence is filtered through slit masks placed behind corresponding microlenses in the detection path. This system enabled 400 Hz volumetric imaging and simultaneous recording from >300 spiking neurons within an ~800-µm-diameter, 180-µm-thick cortical volume (Fig. 3c). With 3D deconvolution, confocal LFM achieved an improved SNR by ~2.4× relative to widefield LFM. More recently, Lu et al. introduced line-confocal illumination in sLFM[58], achieving diffraction-limited, high-speed 3D imaging with SBR performance comparable to spinning-disk confocal microscopy by synchronizing line illumination with the camera rolling shutter, while maintaining a compact system footprint.

## 3. STATE-OF-THE-ART TWO-PHOTON VOLTAGE IMAGING

Compared with 1P excitation microscopy, 2P voltage imaging enables deeper imaging because of reduced optical scattering (Box 2), allowing interrogation of neural dynamics in deep brain regions. However, owing to the much smaller absorption cross-section, 2P voltage imaging faces greater challenges in balancing speed and SNR. In this section, we survey state-of-the-art 2P voltage imaging approaches and classify them into three categories: (*i*) fast scanning, (*ii*) spatiotemporal multiplexing, and (*iii*) scanless holography, each with distinct trade-offs among speed, SNR, and excitation power. Representative performance metrics are summarized in Table 2.

### 3.1 FAST SCANNING

Like 1P confocal microscopy, 2P excitation microscopy typically relies on point scanning to form 2D or 3D images. Conventional raster scanning with galvanometric mirrors, however, is generally limited to only tens of frames per second—far too slow to resolve millisecond-scale membrane voltage dynamics. Faster scanners, including polygon scanners and resonant galvos, can substantially increase throughput. For example, 2P voltage imaging at 440 Hz has been demonstrated in cortical layer 2/3 of awake mice expressing JEDI-2P through a cranial window during visual stimulation,[61] albeit over a highly restricted stripe FOV (e.g., 512 × 32 pixels). This fundamental speed–FOV trade-off can be partially alleviated using an optical scan multiplier unit,[62] which passively rescans a single line using a lenslet array and a retroreflecting mirror to increase effective scan throughput. By combining an 8 kHz resonant scanner with a 16-element lenslet array, this strategy enabled 1,000 fps 2D imaging over a 200 × 200 µm² FOV.[63]

Beyond simply increasing scan speed, throughput can also be improved by reducing the number of samples required per frame. One route is to trade spatial resolution for speed by laterally expanding the excitation spot to approximate neuronal soma dimensions, while maintaining axial confinement via temporal focusing.[64] By reducing the number of scan positions—and therefore the effective pixel count—this strategy enables sampling of thousands of neurons over large volumes (up to $500 \times 500 \times 500$ µm³) at frame rates of ~3–160 Hz. Although promising for large-scale functional imaging, it has not yet been demonstrated with voltage indicators.

A more aggressive reduction in sampling abandons continuous rasterization in favor of targeted readout. For instance, acousto-optic deflectors (AODs) enable random-access targeting by repositioning the focal spot within microseconds, allowing kilohertz-rate sampling at preselected locations. Using this strategy, spontaneous activity in multiple layer 2/3 somatosensory cortex neurons expressing ASAP3 has been monitored in brain slices at 2 kHz.[65] Random-access AOD scanning has also been used to track voltage spikes at 20 discrete points along dendritic arbors in organotypic slice cultures[2] (Fig. 4a) and to perform *in vivo* voltage imaging in layer 5 cortical and hippocampal neurons of awake, behaving mice at sampling rates up to 15 kHz per neuron.[66]

These performance gains come with limitations in scalability and robustness. The limited bandwidth of AODs constrains how many targets can be addressed while maintaining the high sampling rates required for voltage imaging, limiting scalability to large neuronal populations. Emerging alternatives—such as one-dimensional phase modulators[67]—and ongoing developments including AOD-based holography[68] may help expand effective bandwidth. In addition, because random-access strategies acquire relatively little spatial context, they are more susceptible to motion artifacts in freely behaving animals, where movement cannot be retrospectively corrected via image registration. Potential remedies include 3D ribbon scanning[69] and integration of real-time motion tracking.[70]

In parallel, reducing the average optical power required for high-speed 2P voltage imaging is a central objective, both to mitigate heating and to improve photostability. For most currently available voltage indicators, 2P excitation typically requires ~10,000-fold higher average laser power per cell to achieve photon counts comparable to 1P excitation.[71] One strategy to reduce the required average power is to illuminate only when the scan traverses predefined regions of interest (ROIs)—for example, neuronal somata or membranes—analogous to targeted illumination in widefield microscopy.[72] In principle, such temporal gating could be implemented by modulating the laser with an external electro-optic modulator (EOM); however, because gating would occur after amplification, maintaining sufficient pulse energy during the "on" windows can demand impractically high laser power. To overcome this limitation, Li et al. developed an adaptive excitation source (AES) that integrates an internal EOM to gate the 2P seed laser prior to amplification, with a feedback loop to stabilize output.[73] By concentrating excitation only on ROIs during scanning, AES reduces overall sample exposure by more than an order of magnitude.[74]

Another related strategy for high-speed raster imaging is the optical gearbox, which redistributes sampling density to better match the spatiotemporal structure of the specimen. Lin et al. demonstrated imaging speeds of up to 1 kHz using this concept, highlighting its potential as an alternative route to increasing scan efficiency.[75] Although it has not yet been applied to voltage imaging, the optical gearbox represents an intriguing framework for extending fast-scanning microscopy to neural voltage measurements.

Finally, even in the absence of scanner or laser constraints, the maximum pixel rate in raster-scanning 2P microscopy is fundamentally limited by the fluorescence lifetime—typically several nanoseconds—which sets the minimum pixel dwell time needed to avoid temporal crosstalk between adjacent pixels.[76] This lifetime-imposed dwell-time floor restricts the total pixel budget per frame (or volume) at kilohertz acquisition rates, thereby forcing a trade-off between speed and spatial sampling density and ultimately reducing spatial resolution over large 2D or 3D FOVs.

## 3.2 SPATIOTEMPORAL MULTIPLEXING

Beyond increasing scan speed, 2P imaging throughput can be boosted by spatiotemporal multiplexing, in which multiple excitation sites are interrogated per unit time. In the spatial domain, one implementation is parallel scanning, analogous to spinning-disk confocal microscopy: an array of excitation foci is rastered across the FOV while fluorescence from multiple spots is detected in parallel using a detector array. Using this strategy, Zhang et al. generated a $20 \times 20$ focus array with a microlens array and achieved 2P imaging at 1,000 frames per second using sCMOS-based detection.[77]

A conceptually different form of spatial multiplexing replaces point-wise sampling with projection-based (tomographic) measurements that encode signals across the full FOV. When combined with compressive sensing principles, such approaches can substantially reduce the number of measurements required for accurate reconstruction. Kazemipour et al. demonstrated this concept with scanned line angular projection microscopy (SLAP), in which four angular projections are acquired using only four line scans and the image is reconstructed using a simple filtered back-projection algorithm (Fig.

4b).[78] Its ability to record electrically evoked action potentials was demonstrated in hippocampal cultures labeled with the voltage-sensitive dye RhoVR.Pip.Sulf.[79]

Multiplexing can also be encoded in time, enabling single-detector readout without a detector array. Wu et al. introduced free-space angular-chirp-enhanced delay (FACED), a module comprising a pair of slightly misaligned mirrors that converts a single laser beam into a one-dimensional (1D) array of beamlets (Fig. 4c).[80] Using a 920 nm laser operating at 1 MHz, FACED generated 80 excitation foci spanning ~50 µm, with adjacent foci separated by a precisely defined 2 ns temporal delay. This timing offset enables fluorescence from each focus to be read out sequentially using a single photodetector. By sweeping the 1D beamlet array along the orthogonal axis with a galvanometric mirror, the FACED 2P microscope achieved full-frame 2D imaging at 800 frames per second.[81] The system recorded both spikes and subthreshold oscillations from neurons expressing ASAP3-Kv and JEDI-2P-Kv in the visual cortex of head-fixed awake mice at depths up to 345 µm during visual stimulation. More recently, Liu et al. extended the FACED principle using a diffractive beam splitter to generate a 16-beamlet 1D array.[82] When integrated into a standard resonant-galvo 2P microscope, this module expanded the effective FOV by 16-fold and enabled voltage imaging from >260 neurons in the mouse hippocampus.

Finally, multiplexing schemes can combine temporal and spatial encoding with parallel detection. Platisa et al. generated eight excitation beamlets from a single laser source using combined temporal and spatial multiplexing. First, four beamlets were produced using delay lines, with adjacent beamlets separated by 8 ns in time and ~50 µm in space at the sample. These four beamlets were then duplicated via beam splitting and arranged into two parallel rows, yielding eight beamlets spanning ~400 µm along one lateral axis. Fluorescence excited by the beamlets was detected in parallel using a linear photomultiplier tube (PMT) array. This approach, termed SMURF, enabled kilohertz-rate scanning over a 400 × 400 µm² imaging area.[83] Despite their impressive throughput, these multiplexed architectures share practical limitations, including increased system complexity, reduced flexibility in wavelength and speed tuning (as well as FOV adjustment), and sensitivity to pulse-energy stability.

### 3.3 SCANLESS HOLOGRAPHY

In contrast to scanning-based strategies, scanless 2P imaging uses computer-generated holography in combination with temporal focusing to simultaneously deliver excitation to multiple predefined sites (e.g., neuronal somata or membranes) across the FOV (Fig. 4d). By eliminating raster scanning, this approach maximizes excitation dwell time at each target and can improve SNR.[84] Current implementations commonly pair holographic excitation with CMOS-based detection, which offers high sensitivity and substantially higher acquisition rates than conventional PMT readout, albeit with increased vulnerability to tissue scattering.

Sims et al. applied this framework to voltage imaging, including in combination with optogenetic stimulation, and demonstrated *in vivo* multicellular recordings of JEDI-2P signals over a 170 × 300 µm² FOV at depths up to 250 µm.[85] Notably, scanless 2P approaches appear particularly well matched to emerging VSD- and rhodopsin-based voltage indicators[86], positioning them as a promising direction for large-scale, high-speed voltage imaging.

## 4. OPEN CHALLENGES AND OPPORTUNITIES

### 4.1 DEEP 1P VOLTAGE IMAGING

Deep brain imaging is essential for understanding neuronal activity in areas like the hippocampus, crucial for memory and learning. One approach to increase the imaging depth, whilst maintaining the brain intact, is to utilize longer wavelength excitation or emission light, as these wavelengths experience reduced scattering and absorption in biological tissues. However, voltage imaging in the NIR region (NIR-I: ~700–900 nm; NIR-II: ~1000–1700 nm) still faces substantial challenges, primarily due to the scarcity of effective voltage indicators in these longer wavelength ranges. Compared to well-established visible-range indicators, only a few robust, high-performance NIR-I or NIR-II voltage indicators have been developed.[87] This limitation stems from fundamental photophysical constraints. For organic voltage-sensitive dyes, achieving both voltage sensitivity and efficient absorption/emission in the NIR range is complex, as it requires specific properties such as solvatochromism or electrochromism, which are inherently difficult to optimize.[88] Similarly, for GEVIs, shifting spectral properties into the NIR range presents significant challenges since most naturally derived fluorescent proteins, such as GFP derivatives, have emission peaks in the visible spectrum. Recent efforts, such as the development of Archon1[89] and nirButterfly[90], have shown progress in extending voltage indicators into the NIR-I range, but these indicators are still in the early stages of development and lack the maturity and widespread adoption of their visible-range counterparts.[91]

Another critical challenge is the performance limitation of detectors suitable for imaging at longer wavelengths. For example, current camera sensors designed for NIR-II imaging, such as Indium Gallium Arsenide (InGaAs) cameras, typically suffer from low pixel counts due to high fabrication costs and technological complexity. Additionally, these cameras exhibit considerably higher readout and dark current noise compared to traditional silicon-based CCD or CMOS detectors, significantly reducing SNR.[92] This inherent detector noise renders current InGaAs sensors inadequate for the demanding low-light, high-speed requirements of voltage imaging.

Overcoming these obstacles requires a concerted effort to develop both new voltage-sensitive indicators and highly sensitive camera sensors specifically optimized for NIR wavelengths. As an intermediate solution, innovative sampling schemes can be adopted to bridge the gap between the demands of voltage imaging and the current limitations of existing camera sensors. For example, compressive sensing approaches, such as squeezed light-field microscopy demonstrated by Wang et al., enable a substantial reduction in the number of camera pixels required for 3D imaging by capturing only essential neuronal event information.[53] This approach not only increases the camera frame rate but also reduces overall readout noise, as fewer camera pixels are utilized, thereby improving the SNR. Additionally, incorporating confocal detection methods to suppress background fluorescence presents another promising strategy, as exemplified by recent advancements in confocal light-field microscopy.[56] The integration of confocal strategies with NIR voltage imaging could significantly enhance image contrast, ultimately extending the reach of 1P voltage imaging into deeper brain tissues.

The majority of 1P voltage imaging with single cell resolution *in vivo*, including confocal strategies, rely on the detection of ballistic or quasi-ballistic photons from sparsely labelled neuronal populations. This is since, in the case of dense labelling, scattered fluorescence results in broad, overlapping, spatial footprints, and "crosstalk" which makes it difficult to unambiguously attribute activity patterns to individual neurons. A large research effort is currently underway to optimize spatiotemporal de-mixing approaches capable of restoring single neuron resolution. For instance, "activity localization imaging", an approach inspired by single molecule localization microscopy, exploits the relative sparsity of action potentials to identify the spatial footprints of distinct neurons and hence image beyond the ballistic scattering limit.[93] Other approaches aim to operate in diffusive scattering regimes, using the contrast generated by speckle patterns to de-mix temporally varying activity patterns.[94,95] Although these approaches have only been demonstrated in the context of *in vivo* 1P calcium imaging[96], they hold great potential for increasing the imaging depth of camera based 1P and 2P voltage imaging approaches.

## 4.2 PHYSIOLOGICALLY TOLERABLE 2P VOLTAGE IMAGING

As outlined in Section 3, the performance of 2P voltage imaging can be improved along several key axes, including imaging depth, accessible volume, spatiotemporal resolution, and the number of neurons recorded simultaneously. At present, progress is constrained primarily by the maximum excitation power that can be delivered to brain tissue without inducing excessive heating.[97,98] This constraint stems from the substantially higher illumination power per cell required for 2P excitation compared with 1P excitation,[71] and from the need for even higher powers at greater depths to offset reduced peak excitation intensity due to scattering. Together, these factors impose stringent trade-offs between shot-noise–limited sensitivity and photothermal damage. Accordingly, a central priority across state-of-the-art approaches is to reduce the average power required to achieve sufficiently high SNR.

Several clear strategies could help reduce the average excitation power required for 2P voltage imaging while preserving SNR. First, next-generation laser sources that provide high pulse energy with short pulse durations (<70 fs), tunable repetition rates, and excitation wavelengths matched to the absorption peaks of best-in-class voltage indicators could lower linear absorption (and thus heating) while maintaining efficient 2P excitation. Second, because only membrane-localized fluorescence is voltage sensitive, average power could be reduced by restricting excitation to the membrane. This approach, however, would likely require robust real-time motion correction to maintain accurate targeting *in vivo*. A third, comparatively underexplored axis is polarization control: aligning the excitation polarization with the transition dipole orientation can increase 2P excitation efficiency for oriented fluorophores[99,100], and thereby reduce the power needed to reach a given SNR. Finally, continued improvements in voltage indicators themselves—particularly increases in fractional fluorescence changes per unit voltage—directly relax power requirements by boosting signal per photon budget.

Collectively, these advances should accelerate 2P voltage imaging of biologically relevant neuronal ensembles in deep brain regions.

2P voltage imaging at physiologically tolerable power levels would unlock a range of scientific questions that remain largely beyond current experimental reach. A key milestone would be multiplexed recordings from genetically defined neuronal subpopulations, enabling direct comparison of cell-type-specific dynamics with network output and behavior. In addition, crosstalk-free combinations of 2P voltage imaging with optical measurements of neurotransmitter and neuromodulator activity could reveal how synaptic and neuromodulatory inputs shape membrane potential dynamics and inter-neuronal communication. Looking further ahead, all-optical 2P neurophysiology—integrating 2P voltage imaging with 2P optogenetic perturbations—could enable longitudinal mapping of functional connectivity, providing a powerful framework for studying how circuits reorganize during learning, plasticity, and disease progression. Realizing these goals will require continued technical innovation, as well as coordinated efforts to disseminate robust, state-of-the-art 2P voltage imaging platforms to the broader neuroscience community.

## 4.3 SCALING VOLTAGE IMAGING TO MESOSCALE FIELDS OF VIEW

Beyond depth, the next frontier is scaling voltage imaging laterally—from local circuits to mesoscale networks. Voltage imaging over large FOVs is essential because brain function emerges from distributed networks in which distant regions communicate via long-range connections.[101] Capturing these interactions requires mesoscale imaging that can resolve fast voltage dynamics across extended cortical areas. Large-FOV voltage imaging would enable direct visualization of activity propagation, inter-regional coupling, and coordinated population dynamics—providing a critical bridge between cellular electrophysiology and systems-level computations underlying behavior and cognition.

Despite major advances, most voltage imaging demonstrations remain limited to FOVs on the order of hundreds of micrometers.[30,41,56,83] This stands in stark contrast to state-of-the-art calcium imaging, which routinely achieves millimeter-scale mesoscale measurements[102–104] and can monitor activity across multiple cortical regions simultaneously. As a result, the systems-level potential of voltage imaging at cellular resolution remains underexploited, highlighting a pressing need for new approaches that scale voltage imaging to mesoscale dimensions.

For camera-based approaches, including scanless-2P methods, scaling to larger FOVs is often bottlenecked by detector bandwidth: achieving kilohertz frame rates over larger sensor areas requires substantially higher pixel throughput. Encouragingly, camera performance has improved rapidly and is likely to continue doing so. In parallel, architectural strategies such as camera arrays[103,105] and computational strategies such as compressive sensing[53,78,106] offer additional routes to increase effective throughput for large-FOV recordings. A second constraint is the photon budget: high-speed voltage imaging is typically shot-noise limited, making photon collection efficiency paramount and motivating the use of high-NA mesoscope objectives. Recent efforts such as the Cousa objective[107] and the RUSH platform[103] extend FOVs but were primarily optimized for calcium imaging and typically operate at $NA < 0.5$. For voltage imaging, where preserving photons is critical, there is a need for objective designs that simultaneously deliver high NA (to maintain SNR), substantially expanded FOV, and compatibility with *in vivo* preparations.

For scanning-based approaches—particularly 2P microscopy—mesoscale voltage imaging is primarily limited by the throughput of scanning mirrors. Increasing scan speed typically requires smaller-aperture scanners and/or reduced scan angles, which in turn constrains optical resolution and achievable FOV.[62,108] Methods such as optical scan multiplication[62] and spatiotemporal multiplexing[83] can boost effective scan throughput beyond conventional mirror limits. In addition, multi-region imaging[109,110] provides a practical strategy for interrogating long-range interactions without requiring continuous coverage of a full mesoscale area. Inertia-free scanning concepts, including reverberation microscopy[111] and the FACED,[80] further promise to increase throughput by circumventing mechanical constraints. Finally, techniques such as wavefront coding[112] and stereoscopy[113] may offer paths to extend fast 2D measurements into volumetric recordings. Nevertheless, no current method yet provides kilohertz-rate 2P voltage imaging at true mesoscale coverage, and substantial technological development remains necessary.

## 4.4 FROM HEAD-FIXED TO FREELY MOVING: MINIATURIZATION

Finally, realizing network-scale voltage imaging during naturalistic behavior will require not only larger FOVs on the benchtop, but also the ability to shrink these capabilities into head-mounted form factors. Most current voltage imaging systems rely on advanced, highly sensitive, and high-speed camera sensors, as well as powerful excitation lasers. Such systems can accommodate studies involving freely behaving mice only in head-restrained configurations, wherein the animal's head is fixed relative to the imaging objective while still allowing natural limb and body movements. Despite this partial freedom, head-restrained setups inherently limit the range of behaviors that can be studied, restricting access to more naturalistic activities and complex behavioral patterns critical for comprehensive neuroscience investigations. In contrast, calcium imaging has successfully transitioned from conventional benchtop systems to miniaturized, head-mounted microscopes, allowing mice complete freedom of movement during imaging.[114,115] This significant advancement has greatly expanded the scope of behavioral neuroscience studies, enabling detailed investigations of neural activity during complex behaviors, social interactions, and navigation tasks.

For widefield or targeted-excitation voltage imaging, miniaturization is particularly challenging because it requires camera sensors that combine high frame rate with high sensitivity and low noise. Scientific-grade CMOS/CCD detectors typically remain too large and heavy for direct head mounting. Recently, MiniVolt integrated state-of-the-art CMOS sensors into miniature platforms, and demonstrated *in vivo* voltage spike recordings using Voltron2 over a ~250 µm FOV at 530 Hz,[59] but the resulting device is still relatively heavy for freely moving mice (16.4 g). One alternative is to use fiber-optic image bundles to relay fluorescence from the head to a remote detector, enabling the bulky camera to remain off-animal.[116,117] This architecture preserves only essential optics on the head, substantially reducing payload and facilitating multimodal integration without increasing head-borne weight. For example, Szabo et al. combined multi-confocal 1P calcium imaging with targeted optogenetics over a ~250 µm FOV using this strategy.[117] In parallel, recent holographic scanless 2P approaches have begun to overcome the penetration-depth limits of visible-light systems and improve axial confinement, enabling cellular-resolution targeted excitation in freely moving animals over ~250–500 µm FOVs.[118,119] Although commercial fiber bundles typically contain only a few thousand cores—limiting resolution for fine neurites—they can still support soma-level recordings. Extending such relay architectures to voltage imaging, for example by pairing fiber bundles with high-sensitivity cameras (including EMCCD-based implementations already used for scanless voltage imaging in head-restrained preparations), could enable multi-target voltage recordings in freely moving animals, potentially combined with targeted optogenetics and fast calcium imaging.

Scanning-based architectures (e.g., confocal and 2P systems) face a different set of miniaturization constraints. Scanning miniscopes commonly use fiber relays to place the laser source and detector off-animal, leaving only compact imaging optics and a miniature scanner on the head.[120,121] This layout is, in principle, compatible with voltage imaging. However, as in tabletop systems, imaging speed and FOV are ultimately limited by scan throughput, most commonly set by a MEMS mirror. MEMS scanners typically achieve throughputs comparable to resonant galvos and therefore struggle to reach kilohertz imaging without substantial sacrifices in FOV. To address this, strategies demonstrated in benchtop microscopes—such as spatiotemporal multiplexing[122] and multi-region imaging[123]—could be adapted to boost effective throughput in miniaturized scanning-based voltage imaging platforms.

## 5. INTEGRATIVE PERSPECTIVES ON THE TECHNOLOGICAL LANDSCAPE

The recent progress reviewed above makes clear that no single microscopy architecture is universally optimal for voltage imaging. Instead, the best-performing systems emerge from co-design across three tightly coupled layers: the voltage indicator, the imaging instrument, and the downstream inference pipeline (Fig. 5). Voltage indicators determine the achievable signal amplitude, spectral window, membrane localization, photostability, and kinetics; the microscope determines photon throughput, background rejection, spatial coverage, and sampling speed; and the analysis pipeline determines how efficiently weak optical signals can be converted into biologically meaningful readouts such as spikes, subthreshold dynamics, propagation patterns, or functional connectivity. Framing the field through this joint design lens helps clarify why different imaging modalities excel in different regimes and where the next technical opportunities lie.

### 5.1. TASK-DEFINED PERFORMANCE METRICS: SNR, SBR, AND SPATIAL SAMPLING

In voltage imaging, performance should be judged relative to the downstream biological question rather than by frame rate alone. For many experiments, the relevant objective is not simply to maximize raw photon counts, but to maximize the fidelity of a specific inference task - for example, detecting single spikes, resolving subthreshold oscillations, separating neighboring neurons, or tracking activity across depth. Under the shot-noise-limited conditions that characterize many voltage imaging experiments, SNR is fundamentally constrained by photon collection efficiency, making optical throughput a primary design consideration. In this respect, parallel or snapshot detection schemes offer a major advantage because they collect photons from many spatial locations simultaneously rather than sequentially. This "snapshot advantage" is especially valuable when signals are fast, dim, and distributed across a broad field.[124,125]

By contrast, SBR is primarily determined by background rejection. Here, optical-sectioning methods such as confocal microscopy, light-sheet microscopy, and 2P excitation are fundamentally favored because they physically suppress out-of-focus fluorescence before it contributes shot noise to the measurement. Widefield and conventional light-field approaches can partially mitigate background through sparse labeling, targeted illumination, deconvolution, or computational unmixing, but they do not remove the photon noise generated by out-of-focus light that has already reached the detector. This distinction is important: numerical reconstruction can improve contrast in the reconstructed image, but it cannot recover SNR lost to background shot noise at acquisition. Accordingly, one of the most promising design directions is to combine widefield-style parallel detection with physical background suppression, as exemplified by confocal light-field implementations and related hybrid architectures.

Spatial resolution must likewise be interpreted in the context of the intended measurement. For single-cell physiology, the goal may be to resolve soma, dendrites, axons, or conduction dynamics along fine neuronal processes, which pushes microscopy toward high NA, strong optical sectioning, and often volumetric capability. For population imaging, by contrast, the dominant requirement is often reliable demixing of nearby neurons over a larger field and longer duration, even if the system does not preserve submicron detail everywhere. These two regimes place different demands on microscope design and should not be treated as interchangeable use cases.

## 5.2. INDICATOR-INSTRUMENT CO-DESIGN

The properties of voltage indicators should play a more central role in determining imaging architecture. Indicators differ widely in brightness, membrane trafficking, dynamic range, kinetics, photostability, excitation/emission spectra, and susceptibility to background from unresponsive or mislocalized molecules. These properties directly affect whether a modality is limited primarily by photon budget, optical crosstalk, depth penetration, or acquisition bandwidth. For example, dimmer but fast indicators place a premium on optical throughput and may favor parallel detection, whereas indicators with strong membrane confinement and high brightness can better exploit targeted or multiplexed strategies. Likewise, red-shifted or near-infrared indicators could expand opportunities for deeper 1P imaging or reduced scattering, but only if detector sensitivity, optical coatings, and excitation sources are co-optimized accordingly. Thus, microscope design should not be viewed as independent of indicator chemistry; rather, indicator development and instrument development should proceed as a coupled optimization problem.

This co-design principle is especially important in 2P voltage imaging. Because 2P excitation is intrinsically photon inefficient compared with 1P excitation, successful systems often rely on strategies that reduce the number of sampled spatial points, increase dwell time at relevant locations, or multiplex excitation in space and time. These choices are only worthwhile if they remain compatible with the photophysics and localization properties of the chosen indicator. Similarly, in 1P imaging, targeted illumination is most beneficial when labeling density, membrane specificity, and structural priors make it possible to illuminate only informative regions. In both cases, the practical question is not simply which microscope is fastest, but which microscope is best matched to the signal model defined by the indicator and the biological sample.

## 5.3. CAMERA TECHNOLOGY, DATA BURDEN, AND CO-DESIGN WITH COMPUTATION

For camera-based voltage imaging, detector bandwidth has historically been a major bottleneck, particularly for large-FOV or volumetric measurements. That constraint is now shifting. Modern scientific cameras increasingly offer combinations of high frame rate, large pixel count, and high sensitivity that make kilohertz-class recording feasible for selected imaging geometries. As a result, the more consequential bottleneck in many settings is no longer raw sensor performance alone, but the generation, transfer, storage, and analysis of massive time-resolved datasets. This challenge becomes especially severe for long-duration recordings and for 3D methods, where data scale rapidly in both time and depth.

This shift creates an opportunity for deeper co-design between imaging hardware and computational inference. Rather than treating image formation and analysis as separate stages, future systems can be designed so that the acquisition strategy directly reflects the intended inference task. Compressive measurements, optical segmentation, sparse ROI-matched readout, event-driven sensing, and temporally coded acquisition all represent versions of the same idea: do not measure redundant information if the downstream task does not require it. Voltage imaging is particularly amenable to this philosophy because neural activity is often sparse in space, sparse in time, or structured by anatomical priors. Methods such as projection-based imaging, compressive light-field acquisition, and segmentation-guided recording illustrate how reducing measurement dimensionality at the hardware level can substantially ease the computational and storage burden while preserving biologically relevant information. In this sense, the field may benefit less from ever larger raw datasets and more from task-aware acquisition coupled to model-based or learning-based reconstruction.

## 6. CONCLUSION REMARKS

High-speed fluorescence microscopy, coupled with recent advancements in voltage indicators, has significantly expanded the capacity to study fast neuronal dynamics with high spatial and temporal resolution. While calcium imaging remains a valuable tool, voltage imaging offers a direct and precise measurement of membrane potential fluctuations, enabling a deeper understanding of neural circuit computations.

A missed opportunity in the field is the tendency to frame voltage imaging primarily as a technology for large-scale population recording. While that direction is clearly important, voltage imaging also has tremendous potential for single-cell and subcellular physiology, where the design priorities may be quite different. Imaging back-propagating action potentials, dendritic integration, axonal conduction, or compartment-specific synaptic responses demands high spatial resolution, high SBR, and often 3D coverage over elongated structures. These applications may favor confocal, light-sheet, or 2P approaches, potentially combined with extended-depth-of-field strategies, Bessel beams, or other optical designs that better accommodate the intrinsically 3D geometry of neuronal processes. Closer consideration of these distinct biological needs can help users map specific neuroscience questions onto the most appropriate instrument classes, rather than treating all voltage imaging modalities as competitors in a single performance race.

Similarly, the technological landscape could be broadened to include non-imaging readout formats, such as fiber-based or photometry-like voltage measurements. Although these approaches sacrifice spatial information, they can offer substantial advantages in simplicity, stability, depth accessibility, and compatibility with freely behaving preparations. They may be particularly useful when the scientific question centers on population-average voltage dynamics in a genetically defined cell class rather than on resolving individual neurons. Integrating such modalities into the broader voltage-imaging conversation would help unify currently fragmented application domains and emphasize that the central objective is not imaging per se, but robust optical access to fast membrane potential dynamics across scales and preparations.

Overall, the next phase of the field will likely be defined less by isolated improvements in frame rate, resolution, or FOV and more by system-level co-optimization: matching indicator properties to optical architectures, matching acquisition strategies to inference tasks, and matching instrument specifications to the biological scale of interest. By organizing the field around shared constraints and design principles - rather than only around modality labels - we can better identify which trade-offs are fundamental, which are engineering contingencies, and where the most impactful opportunities for innovation now lie.

## ACKNOWLEDGEMENTS

This work was partially funded by National Institutes of Health (R01HL165318, RF1NS128488, R35GM128761, R01NS136027), the Institut Hospitalo-Universitaire FOReSIGHT (P-2PVolt-IHU-000 and the Agence Nationale de la Recherche (ANR-24-CE-6956, 2P-VISMA), and the European Research Council "HOLOVIS-AdG" ERC2019-ADG-885090.

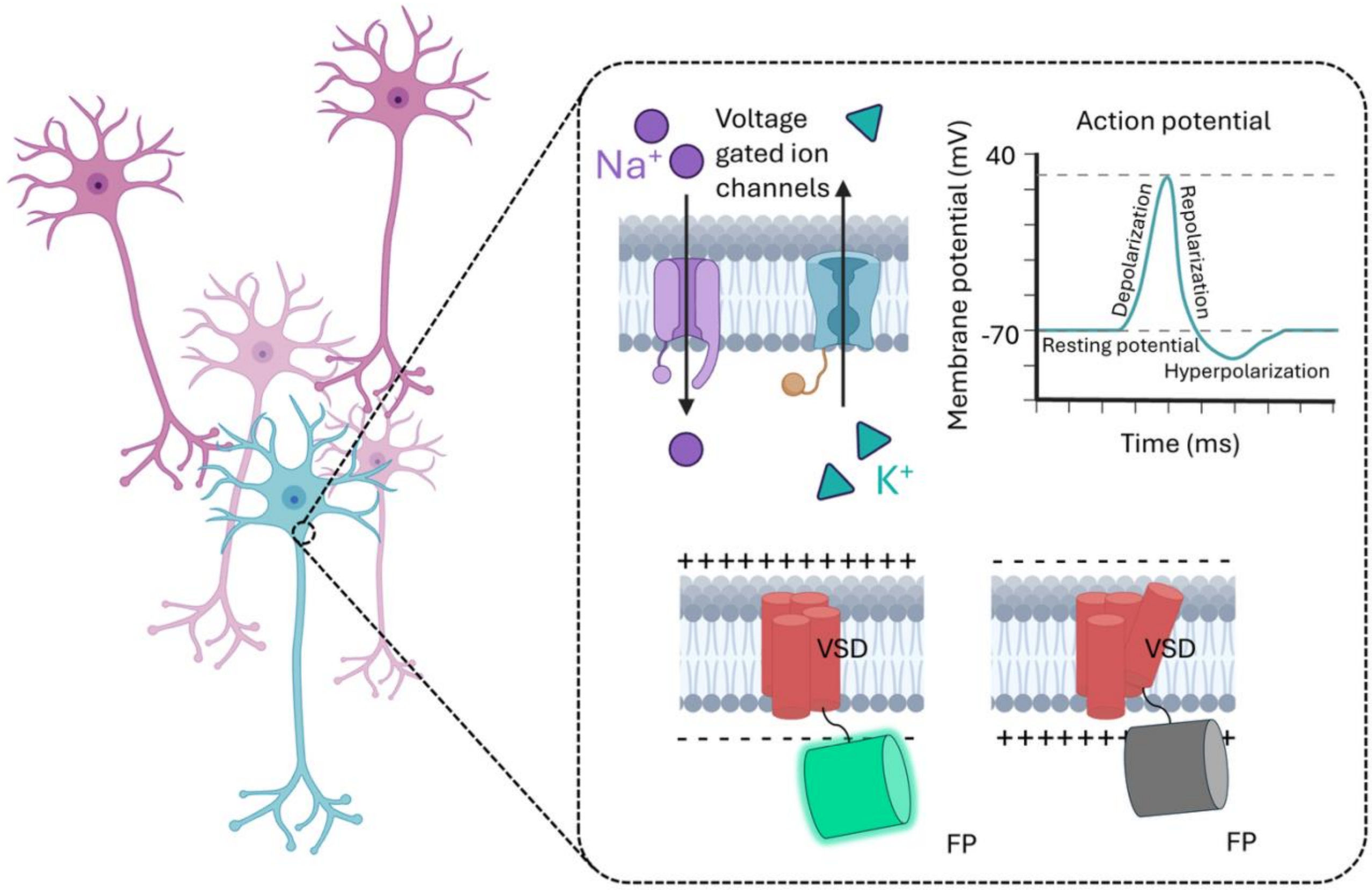

**Box 1 | Membrane Action Potential Signaling in Neurons.** Neurons integrate synaptic inputs across their dendrites and soma. When membrane depolarization reaches threshold, an action potential is initiated by the rapid activation of voltage-gated $Na^+$ channels and subsequently repolarized by voltage-gated $K^+$ channels. This electrical signal propagates along the axon and mediates communication with downstream neurons through chemical neurotransmission or electrical coupling. Modern voltage indicators directly report membrane potential by converting voltage-dependent changes, such as protein conformational shifts or electrochromic effects, into fluorescence signals. As an example, in the ASAP family of GEVIs, a fluorescent protein (FP) is coupled to a voltage-sensing domain (VSD), such that voltage-driven conformational changes in the VSD alter fluorescence intensity.

**Box 2 | Out-of-focus fluorescence versus light scattering**

High-fidelity voltage imaging requires overcoming two distinct optical challenges. Their relative impact depends heavily on the biological preparation and dictates fundamental instrument trade-offs.

- **Out-of-focus fluorescence** arises when fluorophores outside the target focal plane are excited. In densely labeled tissue, this creates a diffuse background that consumes detector dynamic range and degrades the signal-to-background ratio, even in perfectly transparent samples. Light-field microscopy (LFM), for example, lacks intrinsic optical sectioning and is highly vulnerable to this background. To mitigate this, techniques like confocal microscopy and confocal LFM use physical apertures to reject out-of-focus emission , while light-sheet fluorescence microscopy avoids generating it via selective planar excitation. Two-photon (2P) microscopy inherently prevents it through spatially confined, nonlinear excitation.
- **Light scattering** occurs when photons are deflected by refractive index inhomogeneities. Scattering limits excitation penetration and causes scattered emission to form broad, overlapping spatial footprints, making it difficult to unambiguously demix the activity of individual neurons. It is the dominant barrier for deep *in vivo* imaging in mammalian brains, though negligible in transparent organisms like embryonic zebrafish. One-photon techniques rely primarily on ballistic photons, severely limiting depth. Conversely, 2P microscopy utilizes less-scattered near-infrared wavelengths and preserves spatial resolution at depth by assigning scattered emission photons to the scanning focal spot.

**Balancing trade-offs:** Different architectures balance these physical constraints against kilohertz temporal demands. For example, while confocal LFM rejects background, it remains depth-limited by scattering. 2P microscopy elegantly solves both scattering and background challenges but requires complex spatial multiplexing or holographic strategies to achieve necessary speeds without inducing excessive photothermal damage.

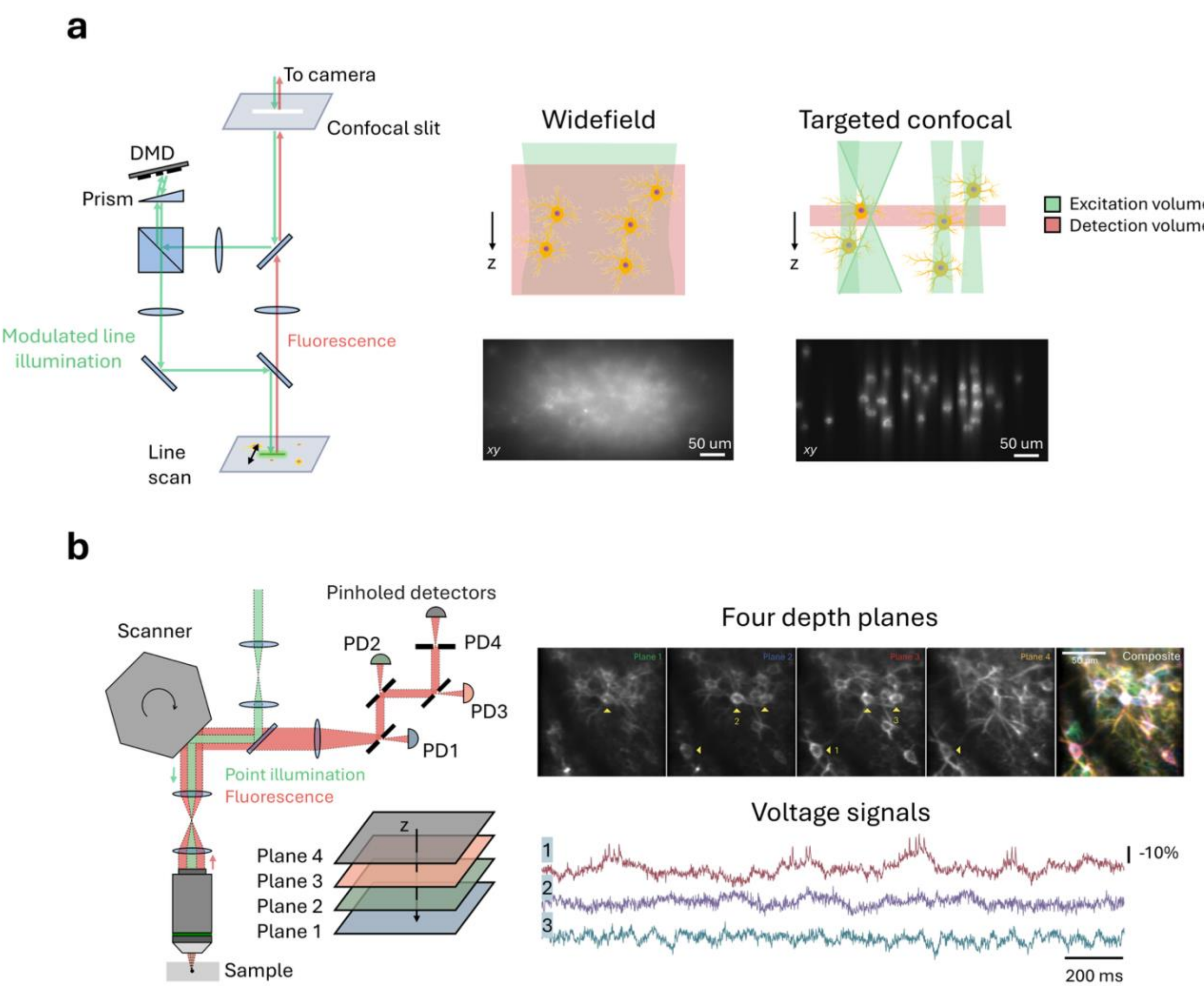

**Figure 1. Confocal approaches for high-contrast voltage imaging.** Confocal detection physically suppresses out-of-focus light, yielding high-contrast neural signals with reduced background noise. **(a)** Combining line-scanning confocal detection with targeted illumination selectively excites only the cells of interest, which minimizes optical cross-contamination and further enhances the signal-to-background ratio. This targeted confocal strategy enables robust *in vivo* voltage recordings in the mouse neocortex. **(b)** Multi-plane confocal microscopy extends this capability into three dimensions by capturing signals from multiple axial depths simultaneously. This enables the concurrent volumetric recording of neuronal activity in the mouse neocortex *in vivo.* DMD, digital micromirror device; PD, photodiode. Figures adapted with permission from ref.[30] and ref.[33] Springer Nature.

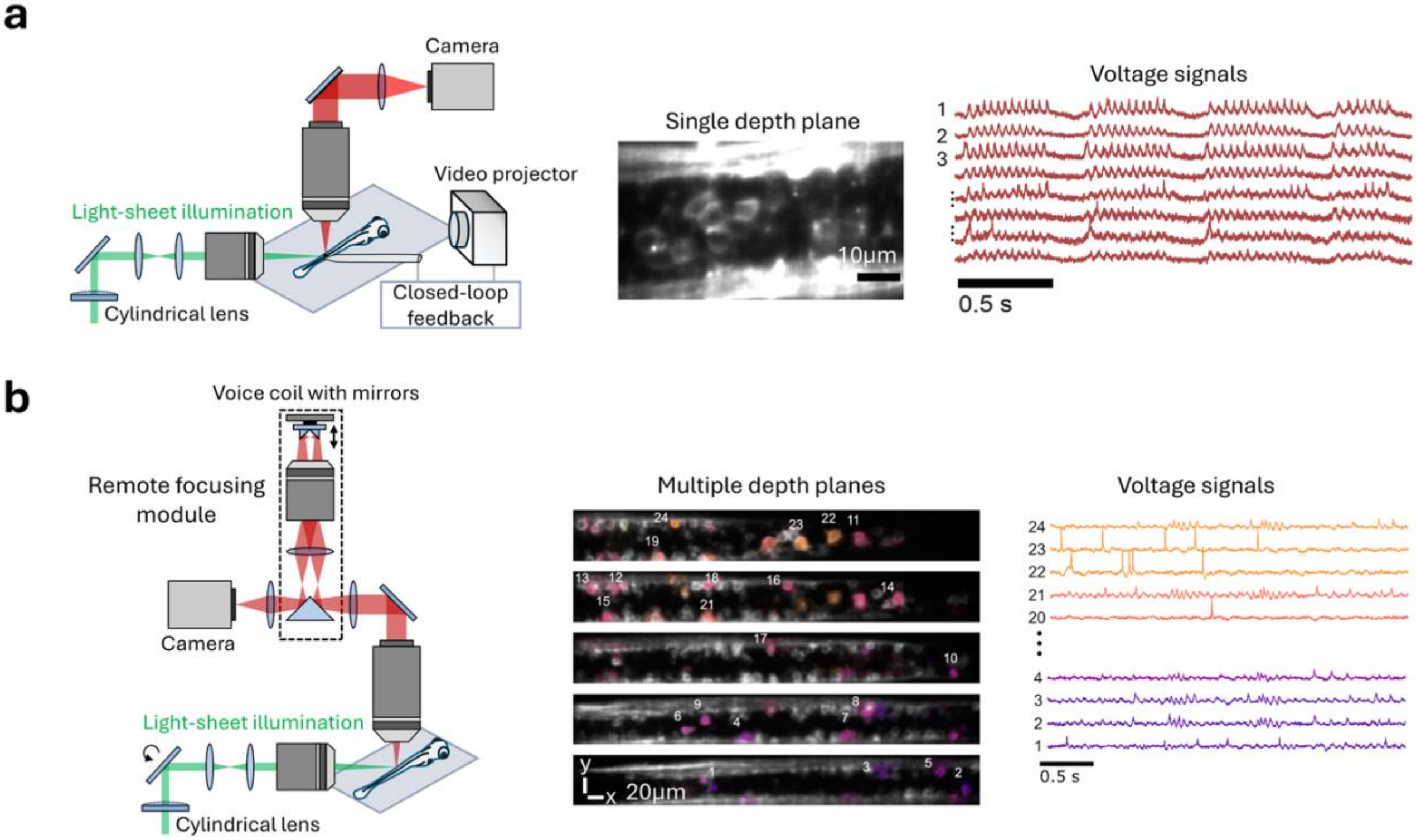

**Figure 2. High-speed light-sheet voltage imaging.** Light-sheet microscopy utilizes selective planar illumination to achieve optical sectioning and rapid acquisition. **(a)** Planar light-sheet setups enable kilohertz-rate voltage imaging, as shown here recording motor activity in larval zebrafish spinal cord during a closed-loop behavioral experiment. **(b)** Integrating a remote focusing module extends this capability to rapid volumetric imaging. By using a fast-moving mirror to shift the focal plane, the system captures 3D voltage dynamics without the mechanical delays of physically translating the primary objective. Figures adapted with permission from ref.[37] and ref.[42].

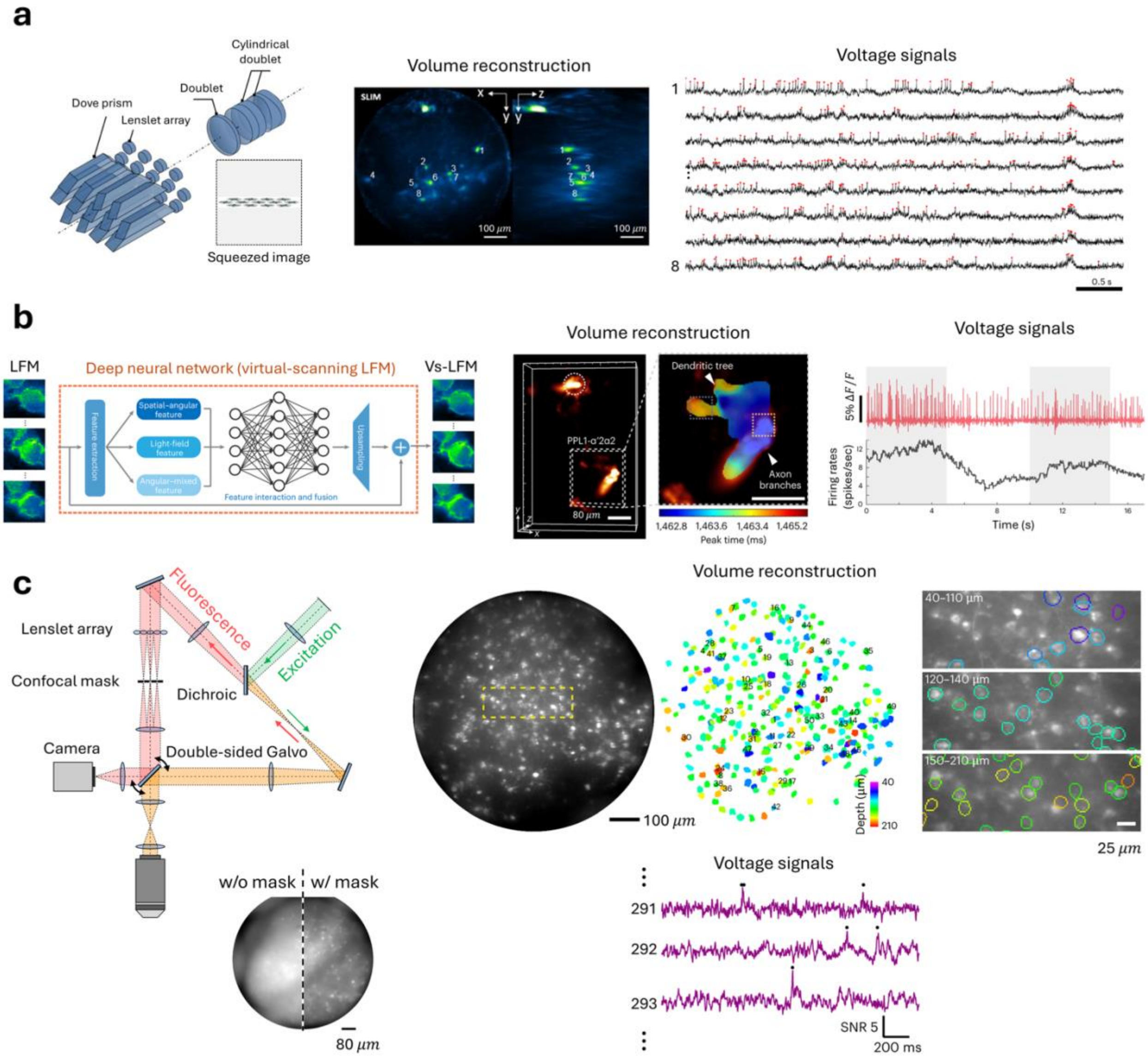

**Figure 3. Snapshot 3D light-field voltage imaging.** Light-field microscopy enables rapid, scanless volumetric imaging by capturing both the spatial and angular information of fluorescent signals in a single camera snapshot. **(a)** Squeezed light-field microscopy (SLIM) utilizes compressive detection to achieve significantly higher volumetric frame rates than conventional setups, enabling kilohertz volumetric imaging of the *in vivo* mouse hippocampus. **(b)** Virtual-scanning light-field microscopy (Vs-LFM) leverages deep learning-based reconstruction to improve spatial resolution without the need for mechanical scanning. This allows ultrafast recording of voltage signals propagating along dendrites and axons, as well as neuronal firing rates, in the adult *Drosophila* brain. **(c)** Confocal light-field microscopy integrates scanning line illumination and confocal masking to actively reject out-of-focus background signals. This improved signal-to-background ratio facilitates high-speed *in vivo* voltage imaging across large neuronal populations (>300) in the mouse cortex. Figures adapted with permission from ref.[55] and ref.[56].

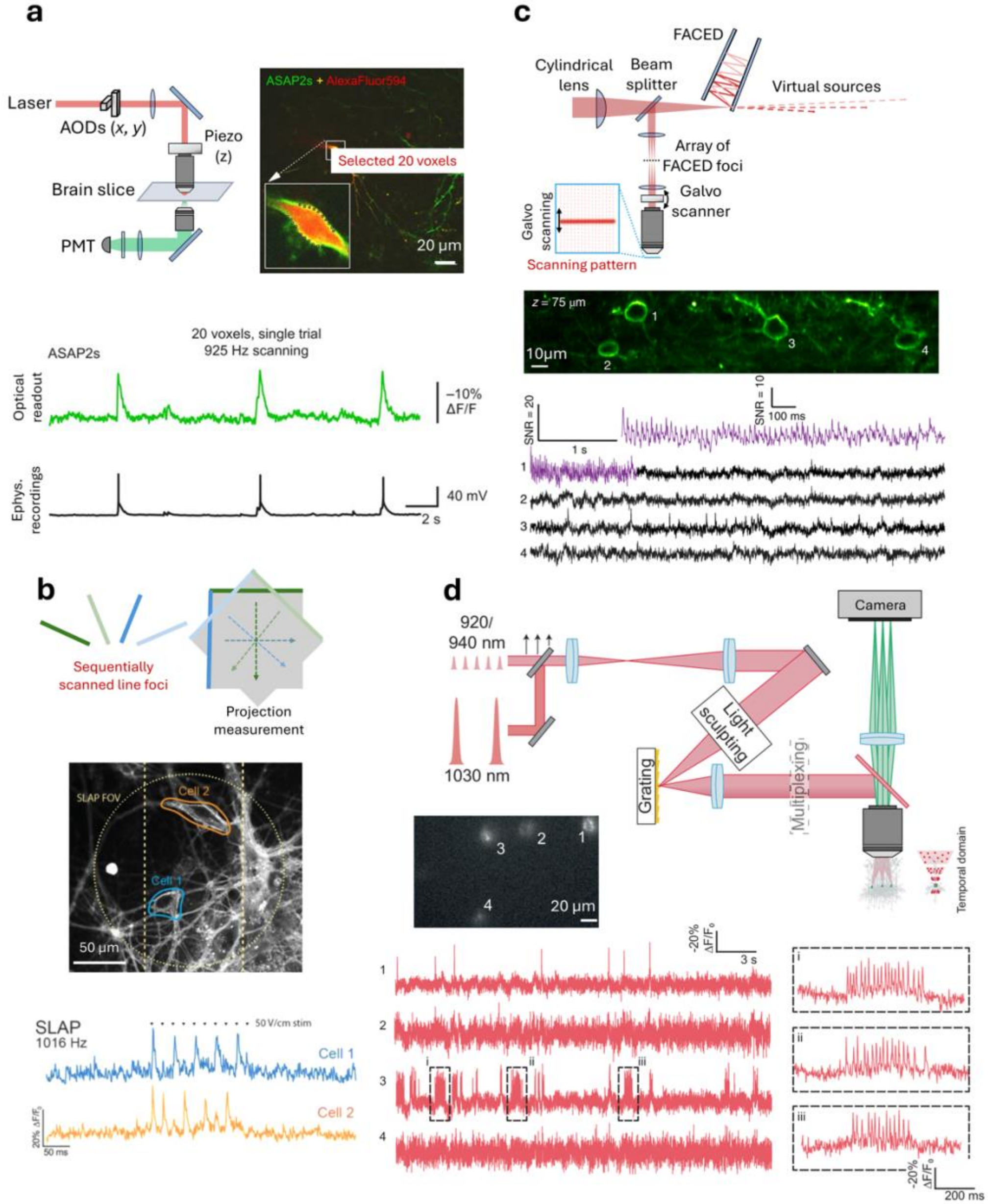

**Figure 4. Strategies for high-speed two-photon (2P) voltage imaging.** Advanced 2P microscopy employs diverse optical strategies to overcome the speed limitations of traditional raster scanning. **(a)** Random-access scanning via acousto-optic deflectors (AODs) precisely targets predefined cellular locations, achieving high temporal resolution by sampling fewer points to image voltage signals in organotypic rat hippocampal slice cultures. **(b)** Spatiotemporal multiplexing via FACED (free-space angular-chirp-enhanced delay) generates an array of temporally staggered beamlets to effectively multiply scan throughput, allowing kilohertz recording of action potentials in the primary visual cortex of an awake mouse. **(c)** Tomographic sampling using scanned line angular projection microscopy (SLAP) relies on scanned angular projections to enable compressive imaging, significantly reducing the required number of measurements to enable kilohertz voltage imaging in hippocampal cultures. **(d)** Scanless 2P imaging simultaneously illuminates multiple targeted regions without mechanical scanning. This maximizes excitation dwell time to boost the signal-to-noise ratio, allowing 500 Hz voltage recording in the barrel cortex of an *in vivo* mouse. PMT, photomultiplier tube. Figures adapted with permission from ref.[66], ref. [80] ref.[78], and ref[85].

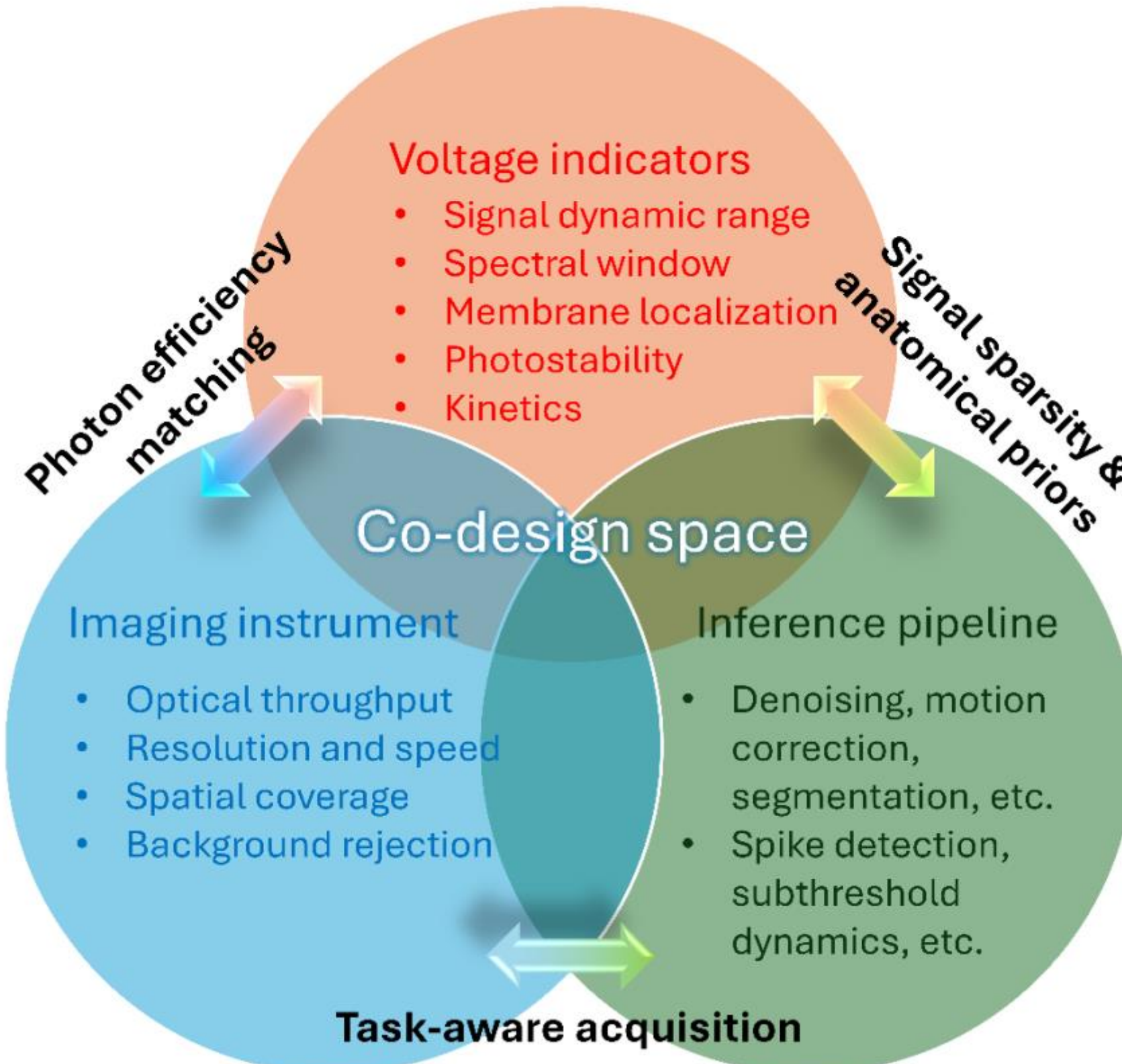

**Figure 5. Integrative co-design framework for voltage imaging.** Optimal performance emerges from the joint optimization of three tightly coupled layers: voltage indicator, imaging instrument, and inference pipeline. The overlapping regions highlight critical intersections, such as matching optical hardware to indicator photophysics and designing task-aware acquisition strategies that directly reflect the intended computational task. Transformative advancements rely on operating within the central co-design space, where molecular properties, optical architectures, and algorithmic pipelines are mutually aligned.

| Technique (Year) | FOV (2D: diameter or X×Y; 3D: diameter × thickness or X×Y×Z) | Frame/ Volume rate (Hz) | Continuous recording duration (best reported) | # Neurons (best reported) | Voltage indicator | Model system applied |
|---|---|---|---|---|---|---|
| **MiniVolt[59] (2026)** | Ø250 μm (2D) | 530 | — | 6 | Voltron2 | Mouse visual cortex V1(L1) |
| **SLIM[53] (2025)** | Ø550×300 μm$^3$ (3D) | 800 | 180 s | 27 | pAce | Mouse hippocampal CA1(SP) |
| **Confocal LFM[56] (2024)** | Ø800×180 μm$^3$ (3D) | 400 | 40 s | 330 | Voltron | Mouse visual cortex V1(L1 and upper portion of L2/3) |
| **FLIPR[60] (2024)** | 390×46×50 μm$^3$ (3D) | 500 | 20 s | >100 | Voltron2 | Zebrafish larvae spinal cord (3 to 4 days post-fertilization). |
| **TICO[30] (2024)** | 1,160 × 325 μm$^2$ (2D) | 800 | 20 min | 78 | Voltron2 & somArchon | Mouse neocortex (L2) and hippocampal CA1(SP) |
| **MuZIC[33] (2023)** | 150×150×45 μm$^3$ (3D) | 916 | — | 40 | Voltron2-ST | Mouse neocortex (L2/3) |
| **Z. Wang et al.[41] (2023)** | 930×370×170 μm$^3$ (3D) | 200.8 | 35 s | 25,556 | Positron2-Kv | Zebrafish larvae whole brain (5 to 6 days post-fertilization) |
| **VsLFM[55] (2023)** | 260×260×100 μm$^3$ (3D) | 500 | 5 s | — | pAce | Adult Drosophila whole brain |

**Table 1. Comparison of representative 1P voltage imaging modalities *in vivo*.**

| Technique (Year) | FOV (X×Y) | Frame rate (Hz) | Continuous recording duration (best reported) | # Neurons (best reported) | Voltage indicator | Model system applied |
|---|---|---|---|---|---|---|
| **FRAME[82] (2025)** | $96\times120\ \mu m^2$ | 1006 | — | Single | JEDI-2P-Kv | Mouse visual cortex V1 (L2/3) |
| | $320\times120\ \mu m^2$ | 805 | 30 s | >200 | JEDI-2P-Kv | Mouse hippocampal CA1 (SP) |
| **FACED 2[81] (2025)** | $160\times400\ \mu m^2$ | 800 | 8.25 s | 182 | JEDI-2P-Kv | Mouse visual cortex V1 (L2/3) |
| | $320\times400\ \mu m^2$ | 385 | 8.32 s | 225 | JEDI-2P-Kv | Mouse visual cortex V1 (L2/3) |
| **AES[77] (2024)** | $31\times365\ \mu m^2$ | 1648 | — | 2 | ASAP5 | Mouse visual cortex V1 (L5/6) |
| | $78\times365\ \mu m^2$ | 701 | — | ≥2 | ASAP5 | Mouse visual cortex V1 (L5/6) |
| **Scanless 2P[85] (2024)** | $86\times300\ \mu m^2$ | 500 | 30 s | 15 | JEDI-2P-Kv | Mouse barrel cortex (L2/3) |
| **SMURF[83] (2023)** | $400\times400\ \mu m^2$ | 803 | Near-continuous for 60 min (59s on, 1s off) | >100 | SpikeyGi2 | Mouse somatosensory cortex (L2/3) |
| **ULoVE[61] (2022)** | — | 440 (resonant scan) | >30 min | 4 | JEDI-2P-Kv | Mouse visual cortex V1 (L2/3) |
| | — | 2500–5000 (random access) | >10 min | 2 | JEDI-2P-Kv | Mouse visual cortex V1 (L2/3, L5) |
| **FACED[80] (2020)** | $50\times250\ \mu m^2$ | 1000 | 1 s | 15 | ASAP3-Kv | Mouse visual cortex V1 (L2/3) |

**Table 2. Comparison of representative 2P voltage imaging modalities *in vivo*.** For techniques demonstrated under multiple experimental configurations, we report each operating regime as a separate sub-entry with matched FOV, acquisition rate, and other parameters.